\documentclass[%
 reprint,
 amsmath,amssymb,
 aps,
]{revtex4-2}

\usepackage{graphicx}
\usepackage{dcolumn}
\usepackage{bm}
\usepackage{subfigure}
\usepackage{graphicx}
\begin{document}

\preprint{APS/123-QED}

\title{{Variation of Gini and Kolkata Indices with Saving Propensity in the Kinetic Exchange Model of Wealth Distribution: An Analytical Study}}

\author{Bijin Joseph}
 \email{bijin.joseph99@gmail.com}
\affiliation{%
 St.\ Xavier's College, Mumbai 400001, India
}%
\author{Bikas K. Chakrabarti}
 \email{bikask.chakrabarti@saha.ac.in}
\affiliation{
 Saha Institute of Nuclear Physics, Kolkata 700064, India}
 \affiliation{SN Bose National Centre for Basic Sciences, Kolkata 700106, India}
 \affiliation{Economic Research Unit, Indian Statistical Institute, Kolkata 700108, India
}%

\begin{abstract}
We study analytically the change in the wealth ($x$) distribution $P(x)$ against saving propensity $\lambda$ in a closed economy, using the Kinetic theory. We estimate the Gini ($g$) and Kolkata ($k)$ indices by deriving (using $P(x)$) the Lorenz function $L(f)$, giving the cumulative fraction $L$ of wealth possessed by fraction $f$ of the people ordered in ascending order of wealth. First, using the exact result for $P(x)$ when $\lambda = 0$ we derive $L(f)$, and from there the index values $g$ and $k$. We then proceed with an approximate gamma distribution form  of $P(x)$ for non-zero values of $\lambda$. Then we derive the results for $g$ and $k$ at $\lambda = 0.25$ and as $\lambda \rightarrow  1$. We note that for $\lambda \rightarrow 1$ the wealth distribution $P(x)$ becomes a Dirac $\delta$-function. Using this and assuming that form for larger values of $\lambda$ we proceed for an approximate estimate for $P(x)$ centered around the most probable wealth (a function of $\lambda$). We utilize this approximate form to evaluate $L(f)$, and using this along with the known analytical expression for $g$, we derive an analytical expression for $k(\lambda)$. These analytical results for $g$ and $k$ at different $\lambda$ are compared with numerical (Monte Carlo) results from the study of the Chakraborti-Chakrabarti model. Next we derive analytically a relation between $g$ and $k$. From the analytical expressions of $g$ and $k$, we proceed for a thermodynamic mapping to show that the former corresponds to entropy and the latter corresponds to the inverse temperature.
\end{abstract}

\maketitle

\section{\label{sec:level1}Introduction}
The economic and financial systems exhibit emerging complex properties of many-body systems and, therefore inevitably, have caught the attention of physicists. Physicists have been studying complex systems for a very long time and have developed several successful models for them. The Kinetic theory, being the oldest and most successful theory of many-body systems, has been developed extensively in the context of income and wealth distribution in societies and applied to markets as a first step \citep{Yakovenko2009,Chakrabarti2013}. \par

In the Kinetic Exchange (KE) model with a uniform saving propensity $\lambda$ of the agents, called the Chakraborti-Chakrabarti (CC) model \cite{Chakraborti2000, Chakrabarti2013}, we take a closed and conserved economic system with $N$ agents having a total amount of wealth, $M$ (here we have taken that $M=N$). These agents are allowed to interact with each other by exchanging their wealth, $x$, through a two-agent money-conserving random stochastic trade process. Here, we have only considered the case where, at a any time $t$, both the agents taking part in any of the binary interactions (trades) are chosen randomly from $N$, and in addition to this, the agents save a constant fraction $\lambda$ (saving propensity) of their wealth, during each exchange (CC model \cite{Chakraborti2000}). In any trade at time $t$, the exchange of wealth $x_i$ of the $i$th with that $x_j$ of the $j$th agent can be expressed as
\begin{equation}
    \begin{split}
        x_i(t+1) =& \lambda x_i(t) + \nu_{i j} ((1-\lambda)(x_i(t) +x_j(t)) \\
        x_j(t+1) =& \lambda x_j(t) + (1-\nu_{i j}) ((1-\lambda)(x_i(t) +x_j(t))
    \end{split} 
\end{equation}
where $0 \le \nu_{i j} \le 1$ is a stochastic fraction varying in every time ($t$) or interaction (trade).  We intend to study how the resulting steady-state wealth distribution $P(x)$ changes with the saving propensity $\lambda$. The inequality in the wealth distribution $P(x)$ can be measured using the Lorenz function \cite{Lorentz} $L(f)$, which gives the cumulative fraction $L$ of wealth possessed by fraction $f$ of the people ordered in ascending order of wealth (explained with greater detail in section \ref{basics}). Using $L(f)$, we can calculate two inequality measures or indices — Gini ($g$) \cite{Gini} and Kolkata ($k$) \cite{ghosh2014inequality, Chakrabarti2020} (also explained with greater detail in section \ref{basics}). Numerical studies have been conducted to determine how the values of indices $g$ and $k$ vary with $\lambda$. In ref. \cite{Paul2021}, a numerical analysis shows that as $\lambda$ increases, $g$ and $k$ decrease (for $\lambda$ = 0 the index value for $g=0.5$ and $k \approx 0.681$, and for $\lambda=1$ the index value for $g=0$ and $k=0.5$). \par
\par
Patriarca, Chakraborti, and Kaski \cite{Patriarca2004} assumed that the steady-state wealth distribution $P(x)$ in the CC model could be represented well by a Gamma function and compared that with the Monte Carlo (MC) results. There have been extensive studies and also applications of this Gamma form for the income or wealth distribution $P(x)$ in the KE models of markets (see for example \cite{Quevedo2020, ribeiro_2020} for important recent applications and discussion on the approximation). In \cite{McDonald}, a general expression for $g$ corresponding to this Gamma approximation of $P(x)$ is given, but the $k$-index still lacks an analytical expression.\par

\begin{figure}[h]

{\includegraphics[width=8.6cm]{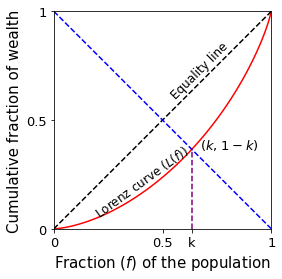}} 

\caption{The Lorenz curve, $L(f)$, represents the amount of wealth possessed by the first $f$ fraction of the population, arranged in increasing order of wealth. The index $g$ is given by the ratio of the area below the Lorenz curve and the area below the equality line. The index $k$ is given by the ordinate value of the intersecting point of the Lorenz curve and the diagonal perpendicular to the equality line. The index value $k$ implies that the $k$ fraction of the poorest people possess only a fraction ($1-k$) of the total income. }
\label{lorentz}
\end{figure}
In this paper, we study analytically the variation of the $g$ and $k$ indices against the saving propensity ($\lambda$) of the agents, using the Gamma distribution form for the wealth distribution $P(x)$ in the CC model. In section \ref{basics}, we introduce the basic quantities we study here. In section \ref{exact}, we  derive the exact values for the $g$ and $k$ indices corresponding to $\lambda=0$ and further, we make use of the Gamma function approximation to analytically calculate the values of the inequality measures corresponding to $\lambda = 0.25$ and $\lambda \rightarrow 1$. In section \ref{assum}, using the results from \cite{Patriarca2004} we make an assumption that for larger values of $\lambda$ the wealth distribution remains as a Dirac $\delta$-function and we proceed for an approximate estimate for $P(x)$ centered around the most probable wealth (a function of $\lambda$). This leads to an analytic form of $L(f)$ and, consequently, of $g$ and $k$ for large values of $\lambda$. These analytical results for $g$ and $k$ at different values of $\lambda$ are compared with the MC results for the CC model. Finally, in this section, we show that initial variation of $k$ with $g$ is linear. In section \ref{sec4}, by integrating over $L(f)$, we expand the Gini index $g(f)$ in powers of $f$ and compare it with the Landau expansion of the free energy functional $F(\eta)$ in terms of the order parameter $\eta$ (see e.g., \cite{landau}). We use this to show that $g$ is equivalent to entropy and $k$ is equivalent to the inverse temperature. Finally in section \ref{summary}, we summarize our work and end with some discussions.
\section{\label{basics} Basic Quantities}
 Inequality in society, in terms of wealth (also of income, votes, citations etc), is ubiquitous. We can measure the inequality in $P(x)$ using the Lorenz function \cite{Lorentz}   
\begin{equation} \label{Lf}
    \begin{split}
       L(f) =& {\int_0^m xP(x)dx \over \int_0^{\infty} xP(x)dx}, \\
       f =& {\int_0^m P(x)dx \over \int_0^{\infty} P(x)dx }.
    \end{split}
\end{equation}
The equality line in Fig. \ref{lorentz} is formed by $L(f)$, when the wealth is distributed equally among all the agents. Using $L(f)$, we can calculate two inequality measures or indices — Gini  ($g$) \cite{Gini} and Kolkata ($k$) \cite{ghosh2014inequality, Chakrabarti2020}. The value of the $g$ index is given by the area between the Lorenz curve and the equality line, normalised by the area ($1/2$) under the equality line:

\begin{equation} \label{gf}
    \begin{split}
       g= 1- 2\left(\int_0^1 L(f)df\right),
    \end{split}
\end{equation} The index $k$ is given by the ordinate value of the intersecting point of the Lorenz curve and the diagonal perpendicular to the equality line. Since each point on the diagonal perpendicular to the equality line will have an abscissa value (say, $k$) equal to the compliment of the ordinate value (correspondingly, $1-k$), the point where this diagonal crosses the Lorenz curve, giving the $k$-index value (see Fig. \ref{lorentz}), has the self-consistent property
 \begin{equation} \label{kf}
   k= 1 - L (k). 
 \end{equation} 
Defining the complimentary Lorenz function $L^c(f) \equiv 1- L(f)$ one can view the $k$-index as its fixed point: $k = L^c(k)$. We can see from Fig. \ref{lorentz} that $k$ fraction of wealth is possessed by $(1-k)$ fraction of the richer population. The values $g=0$, $k=0.5$ corresponds to equality and $g=1$, $k = 1$ corresponds to extreme inequality.
\\
\\
\section{\label{exact} Analytical results from the Gamma function approximation}
In this section we will use the Gamma distribution model to derive the analytical results for $L(f)$ and the inequality indices, corresponding to $n=1$ ($\lambda = 0$), $n=2$ ($\lambda =0.25$) and the limit $n \rightarrow \infty$ ($\lambda \rightarrow 1$). \par
When $\lambda \ge 0$, Patriarca et al. \cite{ Patriarca2004} gives us the following normalized Gamma function representation of $P(x)$ $\left( \int_0^{\infty} P(x)dx = 1\right)$:
\begin{equation} \label{gamma1}
\begin{split}
P_n(x) &= a_nx^{n-1} \exp \left(-{ nx\over \left<x \right>}\right),\\
\end{split}
\end{equation}
with
\begin{equation*} 
\begin{split}
a_n =& {1 \over \Gamma(n)} \left( n \over \left<x \right>\right)^n,\\
\Gamma(n) =& (n-1)!, \\
\end{split}
\end{equation*}
and
\begin{equation} \label{nf}
\begin{split}
n &= 1 + {3 \lambda \over 1 - \lambda}.\\
\end{split}
\end{equation}

\subsection{Case \boldmath $n=1$ ($\lambda = 0$)}
For a vanishing saving propensity ($\lambda=0$), as a consequence of the conservation of money, $P(x)$ relaxes towards an exponential or Gibbs distribution in the steady-state \cite{Chakraborti2000, A2000}. By putting $n=1$ into eqn. (\ref{gamma1})
\begin{equation} \label{p1}
    P(x) = P_1 (x) = {1 \over \left<x \right>} \exp \left(-{x \over \left< x \right>}\right).
\end{equation}
\par
To calculate the $L(f)$, we first find the fraction of the population $f(m)$ having money up to $m$:
\begin{equation*}
    \begin{split}
        f(m) &= \int_0^m P_1(x) dx \\
        &= {1 \over \left<x \right>}  - {1 \over \left<x \right>} \exp\left(-{m \over \left<x \right>} \right)\\
         &= 1- \exp(-m),\\
         \end{split}
\end{equation*}
using $\left<x\right> = 1$ and we get
\begin{equation*}
    \begin{split}
         m &= - \ln(1 - f).
        \end{split}
\end{equation*}
Next we calculate the cumulative wealth $L$ of the population with maximum wealth up to $m$:
\begin{equation*}
    \begin{split} 
        L &= \int_0^m x P_1(x) dx \\
        &= 1 - (m+1) \exp(-m),\\
 \end{split}
\end{equation*}
giving finally (expressing $m$ in terms of $f$)
\begin{equation}
    \begin{split} \label{l1}
        L(f) &= 1 - \left[1- \ln({}1-f)\right](1-f).
    \end{split}
\end{equation}
\begin{figure}[h]
 {\includegraphics[width = 8.6 cm]{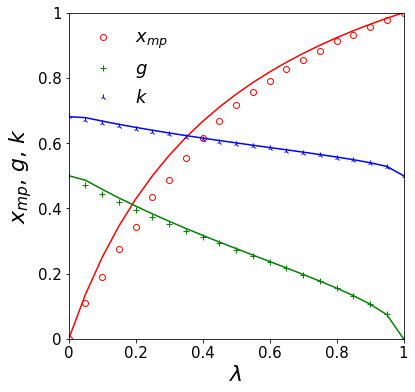}}
 \hfill

\caption{Here we plot the MC simulation results for the variations of $x_{mp}$, $g$-index and $k$-index with $\lambda$ (for averages over the steady-state distribution) in the CC model (shown using markers) and those from Gamma function approximation (eqn. (\ref{gamma1}); shown using continuous lines). In the MC study we took $N=M=500$, and we allowed the system to run for $10^6 - 10^7$ iterations, depending on the requirement to reach the steady-state.} 
\label{fig2}
\end{figure}
\par
From eqn. (\ref{gf}), for $n=1 \ (\lambda = 0$), we get $s = 1/4$, giving $g = 1/2$ and from eqn. (\ref{kf}), we get $k = \left[1- \ln(1-k)\right](1-k)$. From the numerical solution of this self-consistent equation, we get $k \approx 0.682$ for the same case ($n=1$ or $\lambda=0$). \par
As may be seen from Fig. \ref{fig2}, using the MC simulation of the CC model with $\lambda = 0$, we get $g = 0.49 \pm 0.01$ and $k = 0.68 \pm 0.01$ as compared to the analytical estimates here, $g=0.5$ and $k \approx 0.682$ for $n=1$ ($\lambda=0$).
\subsection{Case \boldmath $n=2$ ($\lambda = 0.25$)}
Starting with eqn. (\ref{gamma1}), we consider the case of $\lambda=0.25$ or $n = 2$ and $a_2 = 4$, giving
\begin{equation} \label{p2}
\begin{split}
 P_2(x) & = 4 x\exp(-2x). \\
\end{split}
\end{equation}
Next, we find the fraction $f$ of the population of people having wealth up to $m$;
\begin{equation*}
\begin{split} 
f(m) & = \int_0^m P_2(x) \\
& =  1-e^{-2m}({2m} +1) , \\
m &= {1 \over 2} \left(\left|W_{-1} \left( {f-1 \over e}\right)\right| -1\right).
\end{split}
\end{equation*}
Here $W$ denotes the Lambert $W$-function (see \cite{Corless1996, generalisedW}) and by convention only two branches of $W$, that is $W_0$ (known as the principle branch) and $W_{-1}$, are real valued. However, only for $W_{-1}$ the value of $f(m)$ lies between $0$ and $1$ and therefore we use $W_{-1}$. 
We then find the cumulative fraction of wealth $L$ of people having maximum wealth $m$;
\begin{equation*}
\begin{split} 
 L  =& \int_0^m x P_n(x) \\
  =& 1- e^{-2m} \left({2m^2} + {2m}  + { 1 } \right). \\
\end{split}
\end{equation*}
In the above equation, when $m=0$, $L = 0$ and for $m \rightarrow \infty$, $L=1$.
Replacing $m$ in terms of $f$ we get the Lorenz function
\begin{equation} \label{l2}
    \begin{split} 
 L(f) =&  \left( 1-f \over 2\right) \left[ W_{-1}\left(f-1 \over e\right) + {1 \over W_{-1}\left( f-1 \over e\right)} \right]+1.\\
 \end{split}
\end{equation}
\par
Using thus Lorenz function in eqn \ref{gf} we get, $g = 0.375$, and from eqn \ref{kf} we get,
\begin{equation*}
\begin{split}
1-k =&   \left( 1-k \over 2\right) \left[ W_{-1}\left(k-1 \over e\right) + {1 \over W_{-1}\left( k-1 \over e\right)} \right]+1.\\
\end{split}
\end{equation*}
The numerical solution of this self-consistent equation gives $k \approx 0.634$. \par
As may be seen from Fig. \ref{fig2}, using MC simulation of the CC model with $\lambda = 0.25$, we get $g = 0.37 \pm 0.01$ and $k = 0.63 \pm 0.01$ as compared to the analytical estimates here, $g=0.375$ and $k \approx 0.634$ for $n=2$ ($\lambda=0.25$).
\subsection{Case \boldmath $n \rightarrow\infty$ ($\lambda \rightarrow1$)} 
For $\lambda \rightarrow 1$ both the KE distribution $P(x)$ and its Gamma approximation $P_n(x)$ (with $n\rightarrow \infty$) becomes Dirac $\delta(x-x_0)$ function with $x_0 = 1$.
The fraction $f$ of the population having wealth up to $m$ is given by
\begin{equation*}
 f(m) = \int_0^m \delta(x- x_0)dx = 
    \begin{cases}
      0 & \text{for $m<x_0$} \\
      1 & \text{for $m \ge x_0$ ($=1$)}.
    \end{cases}
\end{equation*}
Next we calculate $L$, the cumulative wealth of the population with maximum wealth up to $m$:
\begin{equation*}
L = \int_0^m x \delta(x-x_0)dx=
    \begin{cases} 
      0 & \text{for $m<x_0$} \\
      x_0 = 1  & \text{for $m \ge x_0$ ($=1$)},
 \end{cases}
\end{equation*}
giving
\begin{equation} \label{l3}
   L(f) = f.
\end{equation}
\par
For $n \rightarrow \infty$ or $\lambda \rightarrow 1$, we get  $g=0$ from eqn. (\ref{gf}). For deriving the $k$-index value we use eqn \ref{gf} and we get $1-k = k$, giving $k = 1/2$ for the same case ($n \rightarrow \infty$ or $\lambda \rightarrow 1$).
\par
As may be seen from Fig. \ref{fig2}, using the MC simulation of the CC model with $\lambda \rightarrow 1$, we get $g = 0.02 \pm 0.02$ and $k = 0.51 \pm 0.02$ as compared to the analytical estimates here, $g=0$ and $k = 0.5$ for $n \rightarrow \infty$ ($\lambda \rightarrow1$).

\section{\label{assum} Results for \boldmath  $\lowercase{g}$ and \lowercase{$k$} by expanding Lorenz function around the most probable value of  wealth}
In this section, we derive analytically the inequality measure $g$ and $k$ for large $\lambda$ values. We know \cite{Chakraborti2000} that as $\lambda \rightarrow1$ (at $\lambda = 1$, the dynamics in the CC model stops) $P(x)$ becomes a Dirac $\delta$-function. Here, we assume that for $\lambda = 1-\epsilon$, where $0<\epsilon<<1$, $P(x)$ remains a Dirac $\delta$-function centered at $x_{mp}$, where $P(x_{mp})$ is maximum (i.e, where ${dP(x) / dx} = 0$). From eqn. (\ref{gamma1}) we can find $x_{mp}$ as follows (using $\left<x \right> = 1$ for $M=N$)
\begin{equation*}
\begin{split}
{d \over dx}P(x) &= {d\over dx} \left(a_n x^{n-1} \exp\left({-nx }\right)\right) = 0. \\
\end{split}
\end{equation*}
This gives
\begin{equation*}
    \begin{split}
  x_{mp} =& {n-1 \over n} = {3 \lambda \over 1 + 2 \lambda}= {3(1-\epsilon)\over 3- 2\epsilon }. 
\end{split}
\end{equation*}
Then the wealth distribution $P(x)$ for $\lambda = 1-\epsilon$, is given by 
\begin{equation} \label{mpi}
    \begin{split}
       P(x) &= \delta(x - x_{mp}), \ x_{mp}= {3(1-\epsilon)\over 3- 2\epsilon }.
\end{split}
\end{equation}
To calculate the $L(f)$, we first find the fraction of the population $f(m)$ having money up to $m$:
\begin{equation*}
     f(m) = \int_0^m \delta(x - x_{mp})dx =
     \begin{cases}
         0 & \text{for $m < x_{mp}$} \\
         1 & \text{for $m \ge x_{mp}$ ($\le1$)}.
    \end{cases}
\end{equation*}
\par
Next we calculate $L$, the cumulative wealth of the population with maximum wealth up to $m$:
\begin{equation*}
 L(m) = \int_0^m x \delta(x-x_{mp})dx =
   \begin{cases}
        0 & \text{for $m < x_{mp}$}\\
        x_{mp} & \text{for $m \ge x_{mp}$ ($\le1$),}
    \end{cases}  
\end{equation*}
giving
\begin{equation*} 
   \begin{split}
       L(f) = x_{mp}  f = {3(1-\epsilon)\over 3- 2\epsilon }  f.\\
   \end{split} 
\end{equation*}
However, here $L(f)$ has only a linear term in $f$. To have minimal non-linearity required for $L(f)$ and to satisfy the requirements $L(0)=0$ and $L(1)=1$ we write, 
\begin{equation} \label{dl}
\begin{split}
    L(f) =& (x_{mp}/A)f + (B/A)f^2; \ A = x_{mp} + B, \\
    x_{mp} =& {3(1-\epsilon)\over 3- 2\epsilon }.
    \end{split}
\end{equation}

Now that we have $L(f)$, we can find $g$. Since,
\begin{equation*}
    \begin{split}
        \int_0^1 L(f) 
         &= {1 \over \left({x_{mp}}  + B\right)}  \left({x_{mp} \over 2}  + {B  \over 3} \right).\\
    \end{split}
\end{equation*}
Hence we get,
\begin{equation} \label{dg}
    \begin{split}
         g &= 1 - {2 \over \left({x_{mp}}  + B  \right)}  \left({x_{mp} \over 2}  + {B \over 3}   \right). \\
    \end{split}
\end{equation}
\begin{figure}[h!]
\mbox{\subfigure{\includegraphics[width=8.6cm]{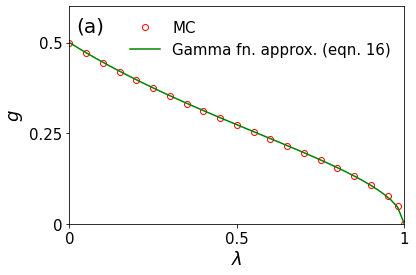}}}
\mbox{
\subfigure{\includegraphics[width=8.6cm]{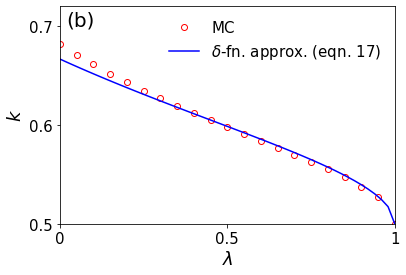}} }
\mbox{
\subfigure{\includegraphics[width=8.6cm]{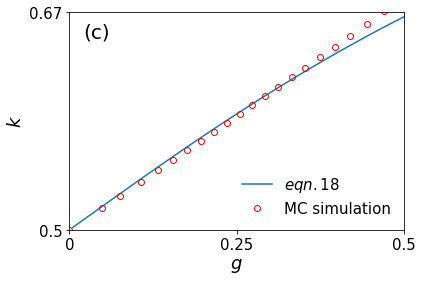}} }
\caption{Results for $g$ and $k$, obtained using (\ref{genG}) and (\ref{gk}), are compared with the numerical solution of the MC results for the CC model. (a) Plot of $g$-index values against saving propensity ($\lambda$) (b) Plot of $k$ against $\lambda$. (c) Gibes plot of $k$ vs $g$, where the initial variations of $k$ vs $g$ is a straight line with slope $3/8$ (as obtained from eqn. (\ref{kl})). From these figures we can see how well the analytical results compare with the numerical data. }
\label{fig3}
\end{figure}
For estimating $k$ we use eqn. (\ref{kf}) giving,
\begin{equation} \label{dk}
    \begin{split}
        Bk^2 + (2x_{mp}+B)k &= (x_{mp}+B).\\
        \\
    \end{split}
\end{equation}
It may be interesting to note that, for $\lambda=0$ eqn. (\ref{dk}) becomes $k^2 + k -1 = 0$, giving the $k$ value equal to $(\sqrt{5}-1)/2$ (inverse of the Golden Ratio). \par
McDonald and Jensen \cite{McDonald} presented an analytic expression for $g$ from Gamma function (\ref{gamma1}) form of $P(x)$: 
\begin{equation} \label{genG}
    \begin{split}
       g = {\Gamma(n+ 0.5) \over \sqrt(\pi) \Gamma(n+1)}. 
    \end{split}
\end{equation}
Here $n$ is given by the eqn. (\ref{nf}). In Fig. \ref{fig3} (a), we have compared this analytical result with the MC result. From eqn. (\ref{dg}) we know that,
\begin{equation*}
    B = \left( 3g \over 1-3g\right) x_{mp}.
\end{equation*}
By substituting this and eqn (\ref{genG}) in eqn. (\ref{dk}), we obtain a general analytical expression for $k$ as
\begin{equation} \label{generalk}
    3\left[{\Gamma(n+ 0.5) \over \sqrt{\pi}
     \Gamma(n+1)}\right]k^2 + \left[2-3 \left({\Gamma(n+ 0.5) \over \sqrt{\pi} \Gamma(n+1)}\right) \right]k - 1 = 0.
\end{equation}
Here $n$, a function of $\lambda$, is given by eqn. (\ref{nf}). In Fig. \ref{fig3} (b), we can see how well this new result agrees with that obtained from the  MC simulation.\par 
In addition to this, by using eqn. (\ref{dk}) and $B$, we derive a relationship between $k$ and $g$
\begin{equation}\label{gk}
    k = {(3g-2) + \sqrt{4+9g^2} \over 6g}.
\end{equation}
We can simplify this non-linear relationship by expanding the square root and neglecting the higher order terms of $g$. This leads to a linear relationship
\begin{equation} \label{kl}
    k= {{1 \over 2} +  {3 \over 8}g },
\end{equation}
for the initial variations of $k$ with $g$. We can see the same in Fig. \ref{fig3}, where the curve appears to be a straight line for smaller values of $g$ and this region has a slope of $3/8$.
\section{Thermodynamic mapping of \boldmath $g$ and
$k$ indices}\label{sec4}
The $\delta$-function approximation of the Gamma distribution $P(m)$ in the CC model discussed in the earlier section can help us to map the $g$ and $k$ indices of the society to the entropy ($S$)  and inverse of temperature ($T$) respectively (also the ordered population fraction $f$, starting from the poorest, as the equivalent order  parameter) of an  equivalent thermodynamic system. \par
As we discussed in the earlier section \ref{assum}, the Lorenz function $L(f)$ in the $\lambda \rightarrow \infty$ limit can be approximated (see eqn.(\ref{dl})) as,
\begin{equation} 
    L(f) = (1/A)[x_{mp}f + Bf^2],
\end{equation}
with $A = x_{mp} +B$, giving $L(0) = 0$ and $L(1) = 1$ as required. From Fig. \ref{lorentz}, one can express the $g$ in  the Landau free energy (entropy) form  (cf. \cite{landau}):
\begin{equation} \label{tg}
    \begin{split}
        g =& 1 - (2/A) \int L(f) df\\
          =& 1 - (2/A)[ (x_{mp}/2)f^2 + (B/3) f^3],
    \end{split}
\end{equation}
with indefinite integral over (arbitrary fluctuation in the order parameter) $f$. We can now compare $g(f)$ with the Landau free energy functional $F(\eta)$ with order parameter $\eta$ (see for e.g. \cite{landau}). It may be noted that as $f$ is positive definite, the $f^3$ term in $g$ as the highest order term in (\ref{tg}) is enough to represent even the continuous (second order) transition in the effective Landau theory for the CC model. \par
Now consider (without any loss of generality) the expression for $g$ (in eqn. (\ref{tg})) up to the quadratic term in  $f$ and compare with that for Landau free energy \cite{landau} $F(T) = (T - T_c) \eta^2$ (as temperature $T$ approaches the transition temperature $T_c$) up to the term quadratic in the arbitrary fluctuation of the order parameter ($\eta$, of which the equilibrium value in the range $0$ to $1$ will be determined by the minimization of $F$) as $g  = 1 - (x_{mp} /A) f^2$. This suggests $exp[(- x_{mp}/A) f^2] \sim exp[(T - T_c)\eta^2]$, giving
\begin{equation} \label{dmp}
    f \sim \eta; A \sim 1/|T -T_c| .
\end{equation}
This indicates that Gini index $g \sim F$, where $F$ denotes entropy  (see e.g. \cite{biro2020gintropy, koutsoyiannis2021entropy}) in the CC model and that the Kolkata index $k$, given by $L(k) = 1 - k$, or $k (x_{mp}/A + 1) = 1$, or $k \sim A \sim 1/|T - T_c|$ for large $|T-T_c|$. Thus $k$ corresponds to the inverse temperature (see e.g., \cite{ghosh2021limiting}). It may also be noted from eqn. (\ref{dmp}) that since $x_{mp} = 3\lambda/(1+2\lambda)$, the transition point $T_c$ occurs in the CC model when the saving propensity $\lambda$ becomes $0$. This is because the fluctuations in the order parameter $f$ here in the CC model is maximum (so are in $g$ and $k$) at $\lambda = 0$ and the fluctuations decrease continuously as $\lambda$ increases, and they vanishes as $\lambda$ approaches unity (where $g$ vanish and $k$ becomes $0.5$ corresponding to the  equality line in  Fig. \ref{lorentz}).
\section{\label{summary}Summary and Discussion}
To summarize, we have studied analytically the variation of the Gini index ($g$) and the Kolkata index ($k$) against the saving propensity ($\lambda$) of the agents, using Gamma distribution \cite{Patriarca2004} ($P_n(x)$ form (\ref{gamma1}) with $n$ from (\ref{nf})) for the wealth distribution $P(x)$ in the Kinetic Exchange Chakraborti-Chakrabarti model \cite{Chakraborti2000}. The notion of wealth in this paper is a generalised one depending on the social context; it is a broad phrase that can be used for money (see e.g., \cite{Chakrabarti2013} for application of CC model to money exchanges), social opinion formation and vote share (see the CC model application to social opinion formation where saving parameter is used as voters' conviction parameter in \cite{vote1, vote2}), citations (see e.g., \cite{kvg1, kvg2} for the observed value $0.37$ of the initial slope for $g$ vs $k$ relationship for citation statistics, derived here analytically in sec. \ref{assum},  eqn. (\ref{kl}) where the slope becomes $3/8$), and so on, and hence the results we obtained can be observed in all of these social contexts.\par 
We start by deriving the exact results for $g$ and $k$ indices corresponding to $\lambda = 0$ and $0.25$, as well as for $\lambda \rightarrow 1$ in sections \ref{exact} and \ref{assum}. This is achieved by assuming the wealth distribution in the CC model to be a  Gamma distribution and using the $\delta$-function approximation of it in the $\lambda \rightarrow1$ limit, and for other values of $\lambda$ the derivation gets involved and requires the use of the generalized Lambert $W$-function \cite{generalisedW}. To avoid this, we concentrate on higher values of 
$\lambda$ ($=1-\epsilon, \ \epsilon \rightarrow0$) and we assume (in section \ref{assum}) the steady wealth distribution to be represented by a $\delta$-function centred around the most probable income ($x_{mp}$) and we find an approximate representation (\ref{dl}) for the Lorenz function $L(f)$.  Using this expression for $L(f)$, we get the analytical results (\ref{dg}) and (\ref{dk}) for $g$ ($\lambda$) and $k$($\lambda$). Using eqn. (\ref{genG}), we obtained the pre-factor $B$, and then compared the MC results for $g$ and $k$ with those from the analytical results (\ref{dg}) and (\ref{dk}) in Fig. \ref{fig3}. Using this method, we were not only able to derive a general expression for $k$, but we also derived a relationship between $k$ and $g$, from which we were able to arrive at the conclusion that the initial variations of $k$ vs $g$ is a linear one with slope $3/8$. This result (initial slope $3/8\approx0.37$) was also already observed in \cite{kvg1, kvg2} from analysis of various datasets, including income distributions of different countries, citation distributions of individual scientists, movie income distribution, votes share distribution among the candidates, etc. It may also be noted from eqn. (\ref{kl}) that for the case of extreme competition (when $g=k$, see e.g., \cite{Chakrabarti2020, ghosh2021limiting}) one gets $k=4/5$, corresponding to the $80-20$ rule of Pareto (see e.g., \cite{Chakrabarti2013}).\par
Further, in section \ref{sec4}, using this $\delta$-function representation of the Gamma distribution (\ref{gamma1}) for $\lambda \rightarrow1$, we obtained and compared the Gini function $g(f)$ in (\ref{tg}) with the Landau free energy functional form $F(\eta)$. We then conclude that the Gini index ($g$) corresponds to free energy ($F$) or entropy (cf. \cite{biro2020gintropy, koutsoyiannis2021entropy}) and the Kolkata index ($k$) corresponds to the inverse temperature of an equivalent thermodynamic system. \par
The Kinetic exchange model, where the agents have saving propensity (as in the CC model \cite{Chakraborti2000}), has been widely studied (see e.g., \cite{Quevedo2020, ribeiro_2020} for recent discussions) and applied with some success in several realistic trade situations. The Gamma function representation \cite{Patriarca2004} of the wealth distribution in the model allows us analytic formulation of the inequality indices like $g$ and $k$ for general values of $\lambda$, as discussed in sections \ref{exact}, and \ref{assum}. The analytical expressions (\ref{dg}) and (\ref{dk}), for $g(\lambda)$ and $k(\lambda)$ respectively, allowed us to get their thermodynamic analogs (discussed in section \ref{sec4}). As we have discussed earlier, many of our results here are already observed in Monte Carlo simulations (for example in \cite{Paul2021}) as well as in data analysis (for example in \cite{biro2020gintropy, koutsoyiannis2021entropy, ghosh2021limiting, kvg1, kvg2}) reported elsewhere.
\section*{Acknowledgments}
We are grateful to Suchismita Banerjee for her comments on the manuscript. We are extremely thankful to an anonymous referee for bringing the reference \cite{McDonald} to our notice. BJ is grateful to the Saha Institute of Nuclear Physics for the award of their Undergraduate Associateship. BKC is thankful to the Indian National Science Academy for their Senior Scientist Research Grant.

\end{document}